\documentclass[review]{elsarticle}
\usepackage{url,hyperref}
\begin{document}

\begin{frontmatter}

\title{What sunspots are whispering about covid-19?} 

\author[TSU]{Mariam M.~Morchiladze}
\ead{Marmorchila@gmail.com}
\author[TSU]{Tamila K.~Silagadze}
\ead{tamilasilagadze7@gmail.com}
\author[NSU]{Zurab K.~Silagadze}
\ead{silagadze@inp.nsk.su}
\address[TSU]{Tbilisi State Medical University, Tbilisi, Georgia}
\address[NSU]{Novosibirsk State University  and Budker Institute of Nuclear
Physics, 630 090, Novosibirsk, Russia.}

\begin{abstract}
Several studies point to the antimicrobial effects of ELF electromagnetic 
fields. Such fields have accompanied life from the very beginning, and it is 
possible that they played a significant role in its emergence and evolution. 
However,  the literature on the biological effects of ELF electromagnetic 
fields is controversial, and we still lack an understanding of the complex 
mechanisms that make such effects, observed in many experiments, possible. 
The Covid-19 pandemic has shown how fragile we are in the face of powerful 
processes operating in the biosphere. We believe that understanding the role 
of ELF electromagnetic fields in regulating the biosphere is important in 
our fight against Covid-19, and research in this direction should be 
intensified.
\end{abstract}


\end{frontmatter}

\section{Introduction}
The covid-2019 pandemic is an extraordinary event in the history of our
civilization.  Billions of people have never been restricted in their homes
before.  Self-isolation is conducive to esoteric reflection, and we
begun to ruminate about Gaia.

Life on earth, once arisen, never completely disappeared, although
it survived at least five major extinctions. Isn't that a small (or big) 
miracle?

Gaia hypothesis suggests that Nature's mercy is neither an accident
nor a benevolent deity's work, but instead is the inevitable result 
of interactions between organisms and their environment \citep{1}.
In poetic terms, the Earth with its ecosystem is a gigantic organism
that harmonizes itself with the help of many invisible feedbacks.
 
Unfortunately, human activity causes environmental degradation,
which in its destructive effect  on the ecosystem  rapidly approaches
the level of the asteroid impact that end the dinosaur era.

We better stop the destabilization of Gaia, because otherwise either
we succeed in this reckless enterprise and destroy the ecosystem and
our own existence, or Gaia recognizes our malicious character and 
invisible feedbacks eliminate us, as a destabilizing element. However,
in order to harmonize with Gaia, we must improve our understanding
of these invisible feedbacks.

\section{Solar activity and pandemics}
At first glance (and at the second too) there can be no connection 
between pandemics and solar activity. However, this is exactly what
Alexander Chizhevsky discovered \citep{2} many years ago. 

Independently, Hope-Simpson  observed the same correlation 
between influenza pandemics and sunspot maximums \citep{3}.
Hoyle and Wickramasinghe confirmed these findings \citep{4}
indicating that the two phenomena have kept in step over some
17 solar cycles.

Interestingly, the previous two Corona virus epidemics, the severe 
acute respiratory syndrome SARS-CoV and the Middle-East
respiratory syndrome MERS-CoV both occurred at double peaks
in the sunspot cycle \citep{5}.
 
A more general result states that most pandemics in the past
occurred near the sunspot extrema (maxima or minima)
\citep{6,7}. In a sense, the present covid-19 pandemic was 
predicted based on this idea \citep{8,8A}.

Of course, such an outlandish idea should be taken with a grain of salt,
and not everybody believes in it \citep{9}. However, we think this strange
idea cannot be dismissed easily. The correlation in the data is so obvious 
that different people have noticed it. We already mentioned  Chizhevsky
and Hope-Simpson. It seems that the first person who linked the 
sunspot cycle to the malaria epidemics, as early as in 1881, was 
C. Meldrum \citep{10}. In 1936, Gill noted that the association of 
malaria epidemics with the epochs of maximum and minimum 
of sunspots is extremely close \citep{10}.

Solar activity can affect biological organisms in various ways. 
Of particular interest is the influence mediated by geomagnetic and
extra-low frequency (ELF) electromagnetic fields \citep{11,12}. 

The Schumann resonance with base frequency of about 8 Hz is a global
electromagnetic resonance excited by natural lightning activity within the
Earth-ionosphere waveguide \citep{13}.  Life evolved on  Earth under the
constant presence of Schumann resonances, and thus, we cannot
exclude that ELF electromagnetic fields played a role in biological 
evolution.

Human activity has become a source of widespread electromagnetic 
pollution, which raises concerns about the possible dangerous health 
effects \citep{14}. Although studies on the possible biological effects of 
ELF and other artificial electromagnetic fields remain largely 
controversial, there is a growing evidence that ELF fields cause
numerous types of changes in cells \citep{15,16,17}.

\section{Biological effects of ELF electromagnetic fields}
Modern life is not conceived without electricity with its power lines and 
appliances, without telecommunication devices. A byproduct of this
technical revolution is an ever-growing number of sources of artificial
electromagnetic fields, both in ELF and radio frequency range, and this
circumstance, as already mentioned, elicits public health concerns 
\citep{14}.

The scope of interactions of electromagnetic fields (EMF) with living matter 
is rich and diverse, requiring simultaneously an unbiased,  open-minded, 
careful and cautious approach when studying the influence of EMF on 
biological processes.

Despite a large body of literature devoted to biological effects of ELF 
magnetic fields (in contrast to  ELF electric fields, magnetic fields easily 
can penetrate biological tissues), no coherent picture has emerged so far 
regarding the plausibility of such effects, or regarding the interaction 
mechanisms.

Epidemiological studies that have focused on the potential health hazards of 
EMFs  are largely controversial. About half of the studies found such effects, 
but the other half failed to find them \citep{18}. The reason for these 
conflicting results is unclear.

Nonetheless,  ample and compelling evidence had been accumulated indicating
that ELF electromagnetic fields has important effects on cell functioning 
\citep{19,20,21,22,23,24,24A}. The nature of these effects is not entirely 
clear. The problem is that such effects are observed for very weak magnetic 
fields, so weak that any such effect is expected to be masked by thermal noise 
\citep{25}.

Perhaps, the extreme sensitivity of living organisms to weak electromagnetic 
fields is not completely unexpected from the point of view of evolutionary 
biology, since life arose and evolved on Earth with the constant presence of 
natural ELF electromagnetic fields,  especially Schumann resonances 
\citep{26,27}.

Remarkably, there is a striking similarity between natural ELF signals and 
human brain electroencephalograms \citep{28}.  Amazingly, many species exhibit, 
irrespective of the size and complexity of their brain, essentially similar 
low-frequency electrical activity \citep{29}, and it is possible that the 
dominant frequencies of brain waves may be an evolutionary result of the 
presence and influence of the resonant ELF electromagnetic background of 
Schumann \citep{30}.

An interesting hypothesis is that ELF background fields played an important 
role in the evolution of biological systems and are used by them as a means of 
stochastic synchronization for various biorhythms \citep{27}. The Schumann 
resonance frequencies are mainly controlled by the Earth's radius, which has 
remained constant over billions of years. Therefore, these frequencies can 
play a special role for the regulatory pathways of living organisms, the 
Schumann resonances providing a synchronization reference signal, a Zeitgeber 
(time giver) \citep{31}.

This hypothesis, while attractive, has a serious drawback: it remains a mystery
how living cells can detect a Schumann resonance signal that is so weak (the 
magnetic component is only several $pT$) compared to the ubiquitous thermal 
noise. The clue maybe is provided by spatially and temporally coherent 
interactions of Schumann resonances with a large ensemble of components of the 
system \citep{32}. For example, the human body contains about $10^{14}$ cells 
and $10^{10}$ cortical neurons.

If the ELF electromagnetic fields, and in particular the Schumann resonances, 
really play a regulatory role in biological processes, then the effect of 
solar activity on living organisms will not look so mysterious, since solar 
activity changes the geomagnetic field and can lead to geomagnetic storms, as 
well as to changes in the parameters of the ionosphere, and, consequently, to 
a change in the parameters of Schumann resonances and ELF radiation  background.

There are some indications that abrupt changes in geomagnetic and solar 
activity, as well as geomagnetic storms, can act as stressors that alter the 
regulatory processes of organisms, blood pressure, immune, neurological, 
cardiac and some other important life-supporting processes in living organisms
\citep{33}. There are studies that indicate that geomagnetic disturbances can 
exacerbate existing diseases, can lead to cardiac arrhythmias, cardiovascular 
disease, a significant increase in hospitalization rates for mental disorders, 
depression, suicide attempts, homicide and traffic accidents (see \citep{33} 
and references therein).

There are several hypotheses that could explain this strange connection 
between solar activity and geomagnetic disturbances and mental health. 
According to the visceral theory of sleep  \citep{34}, the same cortical 
neurons that process exteroceptive information in the waking state switch to 
processing information from various internal organs during sleep. At the same 
time, a violation of the switching process, when visceral information is 
mistakenly interpreted by the brain as exteroceptive,  can manifest itself as 
a mental disorder \citep{35}. If these hypotheses are correct, and if 
geomagnetic disturbances can influence the brain's switching mechanisms, then 
the unexpected link between solar activity and mental disorders could be 
explained.

\section{Concluding remarks}
Scientific progress has greatly expanded the ability of humankind to cause 
large-scale changes in the environment.  Unfortunately, we do not always 
understand the subtle feedback loops that operate in the biosphere to predict 
all consequences of such changes. An amusing example of the unexpected outcome 
of a large-scale human intervention in nature is provided in \citep{36}. 

For unknown reasons, fish in the Gulf of Mexico position themselves over 
buried oil pipelines off the shore of Texas, orienting themselves directly 
above the buried pipeline at a height of 1-3 meters above the seabed and 
perpendicular to the axis of the pipeline. Presumably they are responding to 
some electromagnetic stimuli, such as remnant magnetism in pipeline sections, 
voltage gradient induced by corrosion protection devices, or transient signals 
induced into the pipeline by remote lightnings or solar wind induced magnetic 
storms \citep{36}.

The research of biological effects was intensified at about 1967 as part of an
evaluation of the environmental impact of a proposed ELF military antenna 
(Project Sanguine) \citep{37}. Unfortunately, the presence of military and 
commercial components in this research makes it politically very sensitive 
\citep{26}. Nonetheless, in light of the results so far available, it would 
be too irresponsible to dismiss such effects as being  implausible \citep{32}.

On the practical side, if ELF fields cause biological effects, whatever 
the unknown mechanism of this interaction, we can try to use these effects 
in our fight against SARS-CoV-2 and similar infections. It is known that
bioelectric signals generated by the metabolic activity of cells are in an 
ELF range, therefore by interfering with these signals by external low 
intensity ELF electromagnetic fields we can suppress microbial or
bacterial activities \citep{18a,19a,20a}.

We believe that under the burden of the Covid-19 pandemic, research in 
this direction should be intensified. Studies of the antimicrobial effects 
of ELF electromagnetic fields are not expected to be too expensive. 
If successful,  it promises a non-invasive, inexpensive, safe and fast 
technique to fight infections \citep{19a,20a}.

Is Covid-19 pandemic related to the deepest sunspot  minimum for a century
we are experiencing now? During sunspot  minimum solar magnetic field gets 
weaker and, as a result, galactic cosmic rays flux entering the earth 
increases. There are some indications that an increase in the flux of cosmic 
rays can lead to an increase in lightning activity on earth \citep{38,39} and 
thus change the natural ELF electromagnetic background. As noted in 
\citep{26}, changing the electromagnetic background poses a twofold challenge 
to us: weakening the immune system due to constant stress and more severe 
illnesses, since electromagnetic fields can stimulate bacterial growth and 
increase their resistance to antibiotics.

Increased cosmic rays can lead also to appearance of novel virion strains due 
to induced mutation  and genetic recombination events \citep{8A,40}, 
especially if viruses spread even beyond the tropopause (new bacteria 
have been found in the stratosphere and even on the exterior of the 
International Space Station, orbiting at a altitude of 400~km) \citep{40}.

Interestingly, the idea that the flux of galactic cosmic rays can affect 
the ELF electromagnetic background can be tested using the Forbush effect
\cite{39A}. During solar flares, the flux of galactic cosmic rays decreases
rapidly (over a day or less) due to modification of the near-Earth 
interplanetary magnetic field. This so-called Forbush decrease is transient 
and is followed by a gradual recovery over several days \cite{39B,39C,39D}.

Based on the measurements in the Kola peninsula, it was demonstrated that 
in all ten events of significant Forbush-decreases, the  intensity of the
ELF-atmospherics decreased (down to their complete disappearance) \cite{39A}.
It was hypothesized that this phenomenon is caused by a decrease in the 
intensity of discharges of a special type (sprites and jets) as a result of 
a decrease in atmospheric ionization at altitudes of 10-30~km during the 
Forbush decrease in the flux of galactic cosmic rays \cite{39A}.

Cosmic ray forcing of the climate acts simultaneously and with the same 
sign throughout the entire globe and operates on all time scales from days 
to hundreds of millions of years \cite{39E}. For this reason, even a 
relatively small forcing can lead to a large climatic response over time
\cite{39E}. To unravel the anthropogenic contribution to the current climate 
change and assess its danger, which is now the subject of much public 
concern and controversy, we need to understand physical mechanisms underlying 
the influence of solar and cosmic ray variability on climate and their impact 
on the biosphere.  
 
It has recently been shown that bats, like many other animals with highly 
developed magnetosensory skills, use magnetic field for orientation 
and can sense even very weak magnetic fields \cite{40A,40B,40C}. Perhaps, 
this magnetoreception is influenced by ELF electromagnetic background
\cite{40D,40E}. Another source of possible influence is the change in 
cloudiness due to the increased flux of galactic cosmic rays, as bats have 
been shown to calibrate their magnetic compasses with sunset cues \cite{40F}.  

So, one can imagine the following scenario \footnote{Suggested by an 
anonymous reviewer}. Changes in the ELF electromagnetic background, caused 
by the increased flux of cosmic rays due to unusually deep sunspot minimum, 
can cause abnormal movements of the population of bats and affect the time 
of their arrival and departure. Delayed arrival or departure and longer travel 
times can increase the population of bats in some areas, thereby increasing 
competition for limited food supplies, and can also increase the likelihood 
of interspecies transmission of the virus. Besides, increased level of 
irradiation increases genetic recombination rates, as was demonstrated
in laboratories during the 1950s and 1960s \cite{6}. Finally, under the 
influence of these circumstances, the new coronavirus successfully recombined 
and caused the Covid-19 pandemic.

We end this article with a funny observation from \citep{40}. The Italian word 
influencia means influence, meaning the influence of the stars in the case of 
influenza illness. This etiology reflects the belief of our ancestors that 
events in the sky and events on Earth are interconnected. It may well be that 
they were right.

\section*{Acknowledgments}
The authors thank Olga Chashchina and an anonymous reviewer for critical 
comments that helped to improve the manuscript.

\section*{Conflict of interest statement}
The author declares that the research was conducted in the absence of any 
commercial or financial relationships that could be construed as a potential 
conflict of interest.

\end{document}